\documentstyle[art10,epsf]{article}
\epsfverbosetrue
\def\as{\alpha_{\rm s}}
\def\O{{\cal O}}

\def\lsim{\mathrel{\mathop
  {\hbox{\lower0.5ex\hbox{$\sim$}\kern-0.8em\lower-0.7ex\hbox{$<$}}}}}
\def\gsim{\mathrel{\mathop
  {\hbox{\lower0.5ex\hbox{$\sim$}\kern-0.8em\lower-0.7ex\hbox{$>$}}}}}
\def\to{\rightarrow}

\def\pT{p_{\rm T}}
\def\pbp{p\bar{p}}
\def\epem{e^+e^-}
\def\rs{\sqrt{s}}
\begin{document}
%
\begin{titlepage}
\noindent
DESY 95-020  \hfill  ISSN 0418-9833     \\
TUM-T31-84/95                           \\
January 95\hfill                        \\[9ex]
\begin{center}
{\Large \bf Inclusive particle production
             in $\pbp$ Collisions                    }    \\[10ex]
{\large F.M.\ Borzumati$^{1,2}$, G.\ Kramer$^{1}$    }    \\[4ex]
{\it $^{1}$ II.\ Institut f\"ur Theoretische Physik      \\
 Universit\"at Hamburg, 22761 Hamburg, Germany      }    \\[1.2ex]
{\it $^{2}$ Institut f\"ur Theoretische Physik,          \\
 Technische Universit\"at M\"unchen, 85747 Garching,
                                            Germany }    \\[21ex]
{\large \bf Abstract}
\end{center}
\begin{quotation}
We calculate the inclusive production of charged hadrons
in $\pbp$ collisions to next-to-leading order (NLO) in the QCD
improved parton model using a new set of NLO fragmentation functions
for charged pions and kaons. We predict transverse-momentum
distributions and compare them with experimental data from the CERN
S$\pbp$S Collider and the Fermilab Tevatron.
\end{quotation}
\vfill
DESY 95-020      \hfill      \\
TUM-T31-84/95    \hfill      \\
\end{titlepage}
%
\twocolumn
\section{Introduction}
%
The inclusive production of single hadrons in hadron-hadron,
photon-hadron and deep inelastic lepton-hadron collisions is an
important area to test the QCD improved parton model. The inclusive
cross section is expressed as a convolution of the parton distribution
functions, the partonic cross sections and the fragmentation functions
of quarks and gluons into charged or neutral particles. The
factorization theorem ensures that parton distribution and
fragmentation functions are universal functions and that only the hard
scattering partonic cross sections change when different processes are
considered. This theory provides a rather consistent description of
many large-momentum-transfer-processes\,[1] and it is suitable for
describing the inclusive production of single hadrons at large
transverse momentum ($\pT$) in various reactions. In this work we
study the inclusive production of charged pions and kaons in high
energy $\pbp$ collisions.

Experimental data for charged single particle production come from the
CERN ISR pp Collider\,[2]; from the UA1\,[3] and UA2\,[4]
Collaborations at the S$\pbp$S Collider at CERN; and from the CDF
Collaboration at the Tevatron\,[5]. Recently, high statistics data
from the UA1 MIMI Collaboration\,[6] have become available which has
extended the $\pT$ range to much larger values as compared to earlier
UA1 analysis.

In\,[7], together with B.A.\,Kniehl, we already presented results for
inclusive single-charged-hadron and single-$\pi^0$ cross sections at
the next-to-leading order (NLO)\,[7]. We compared them with
experimental data from the UA2 and CDF Collaborations. The overall
agreement concerning the $\pT$ dependence and the absolute
normalization of the cross sections between theoretical and
experimental results was satisfactory even for the smaller $\pT$
region. An important drawback of this work, however, was due to the
use of leading order (LO) parametrizations of fragmentation functions
for charged pions and kaons\,[8].  These old parametrizations had been
constructed more than ten years ago from fits to the data then
available, obtained in low-energy $\epem$ annihilation experiments and
deep inelastic muon-nucleon scattering.

Recently a NLO set of parametrizations for fragmentation into charged
pions and kaons was obtained\,[9]. These parametrizations are
generated through fits to $\epem$ annihilation data taken at
$\rs=29$\,GeV by the TPC Collaboration. They produce rather
satisfactory fits also for other $\epem$ data for charged particle
production obtained at lower energy at DORIS and at higher energy at
PETRA, PEP and LEP.

By using these new parametrizations, we are now in position to perform
a full NLO calculation including NLO parton distribution functions,
NLO parton-parton hard scattering cross sections and NLO fragmentation
functions thereby, removing a serious limitation of our earlier
work\,[7].

Other authors have almost simultaneously presented NLO
parametrizations of fragmentation functions for light mesons,
i.e. neutral pions\,[10], eta mesons\,[11] and charged pions\,[12]. In
general these parametrizations are obtained through fits to data
produced by the HERWIG Monte Carlo\,[13] at fixed
$Q_0=30$\,GeV. Parametrizations at lower scales are then obtained via
NLO evolution. The authors of\,[10] and\,[11] compared their
$\pi^0$--fragmentation functions with data from pp collisions in fixed
target experiments and the ISR, and from $\pbp$ collisions at the CERN
S$\pbp$S Collider.  In\,[10] also, several sets of fragmentation
functions for $\pi^0$ and $\eta$ were obtained at a low scale,
comparable to the one used in\,[8] through simultaneous fits to
$\epem$ annihilation, fixed target and ISR $pp$ collisions and
S$\pbp$S Collider data.

In this work we follow a different approach. We assume that the NLO
fragmentation function for charged pions and kaons are sufficiently
constrained by the fit to the TPC $\epem$ annihilation data and the
fit to the gluon fragmentation function from the OPAL
measurements\,[14]. Our aim is to verify whether the NLO
parametrization of\,[9] gives also a satisfactory account of existing
$\pbp$ collider data and whether it improves the estimates obtained in
our earlier work\,[7]. As in\,[7], we assume that the charged particle
yield can be very well described by the sum of the pion and kaon yield
and that the inclusive production of other charged particles, like $p$
and $\bar p$ and heavier baryons is negligible compared to the
production of the two lighter mesons.

This work is organized as follows. In Sect. 2 we give a short
introduction of the formalism and fix our inputs. The results of our
calculation and a comparison with data from $\pbp$ colliders are
presented in Sect. 3. This sections ends with a discussion of the
results and concluding remarks.

\section{Formalism and Input}
%
Before presenting the results, we give a brief introduction to the
NLO formalism. The NLO inclusive cross section for the production of
a single hadron $h$ in the reaction
$$
 p(p_1)+\bar{p}(p_2)\to h(p_3)+X\,,
\eqno(1)
$$
 is written as:
$$
E_3 {d^3\sigma \over d^3 p_3} =
\sum_{a,b,c} \int dx_1 \int dx_2 \int{dx_3\over x_3^2}
 \qquad
$$
$$
 \qquad{}\times
F_a^p (x_1,M^2)\, F_b^{\bar p}(x_2,M^2)\,D_c^{h}\left(x_3,M_f^2
\right)
 \qquad
$$
$$
 \qquad{}\times
 {1 \over \pi S } \left[ {1 \over v}
 { d  \sigma_{a b\to c}^0(s,v,\mu^2)\over dv}\, \delta(1-w) \right.
\qquad
$$
$$
 \qquad{} +
  \left.  {\as(\mu^2)\over2\pi}\,
  K_{a b \to c}\left(s,v,w;\mu^2,M^2,M_f^2\right) \right]\,.
\eqno(2)
$$

\begin{figure}[tb]
\epsfxsize=8.30cm
\epsfbox[61 131 534 756]{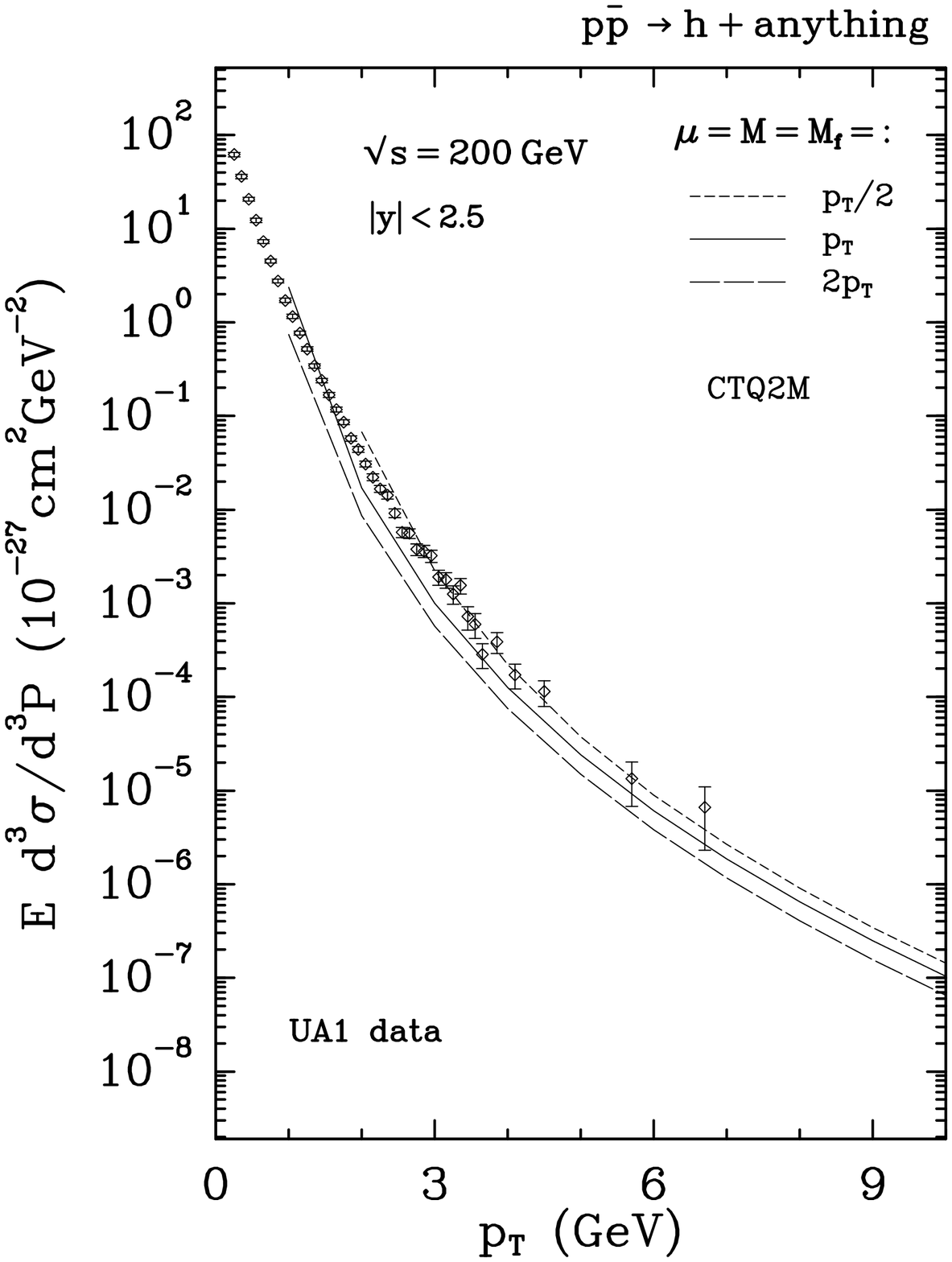}
{{\small {\bf Fig.~1.}
Inclusive cross section for production of
 charged-hadrons ($h\equiv (h^+ +h^-)/2$) as a function of $\pT$
 for $\sqrt{s}=200\,$GeV and rapidity $\vert y \vert < 2.5$. The
 short-dashed, solid and long-dashed lines
 correspond to the full NLO prediction for scales $\mu$, $M$ and $M_f$
 set equal to $\pT/2$, $\pT$ and $2 \pT$. For comparison, the UA1
 data\,[3] taken at the same 
 energy and in the same rapidity range are also shown }}
\end{figure}

The partonic variables $v$ and $w$ are related to the usual $s,t,u$
($s=(p_a+p_b)^2$, $t=(p_a-p_b)^2$ and $u=(p_b-p_c)^2$) as:
$v=1+t/s$, $w=-u/(s+t)$. They are related to the hadronic variables
$S =(p_1+p_2)^2$, $T=(p_1-p_3)^2$ and $U=(p_2-p_3)^2$ by:
$$
 s=x_1x_2S\,,\qquad
 t={x_1\over x_3}\,T\,,\qquad
 u={x_2\over x_3}\,U\,,
\eqno(3)
$$
The indices $a,b,c$ run over gluons and $N_F$ flavours of quarks. As
in our earlier work\,[7], we assume $N_F=4$ and neglect the influence
of the charm-quark threshold.  $F_a^{p}(x_1,M^2)$ and
$D_c^{h}(x_3,M_f^2)$ are the usual
structure and fragmentation functions for partons of
type $a$ and $c$ respectively inside the proton and the
hadron $h$. They depend on $x_1$, $x_3$, the fractions of proton momentum
carried by parton $a$ and the fraction of the momentum of
parton $c$ carried by hadron $h$ and on
the factorization scales $M$ and $M_f$.
The additional mass parameter $\mu$ is the renormalization scale for
the strong coupling.
Finally, $d^3\sigma_{a b \to c}^0$ is the LO
partonic cross section for the process $a+ b\to c+X$ in
$\O\left(\as^2(\mu^2)\right)$. The $K_{a b \to c}$ functions are the
NLO corrections to $a + b \to c + X $ and are taken from the work by
Aversa et al.\,[15].

\begin{figure}[tb]
\epsfxsize=8.30cm
\epsfbox[61 131 534 756]{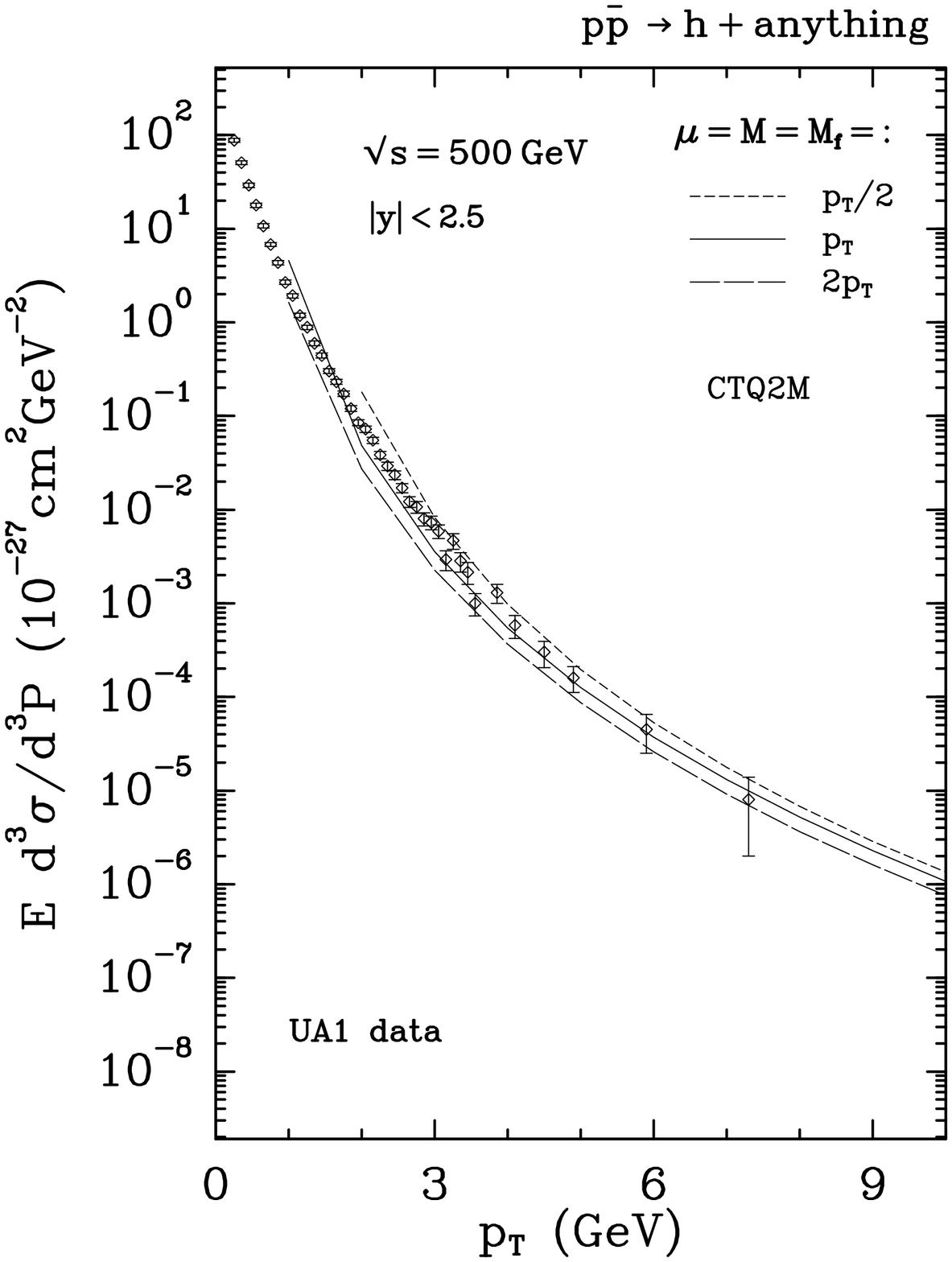}
{{\small {\bf Fig.~2.}
Same as in Fig. 1 for $\sqrt{s}=500\,$GeV }}
\end{figure}
\begin{figure}[tb]
\epsfxsize=8.30cm
\epsfbox[61 131 534 756]{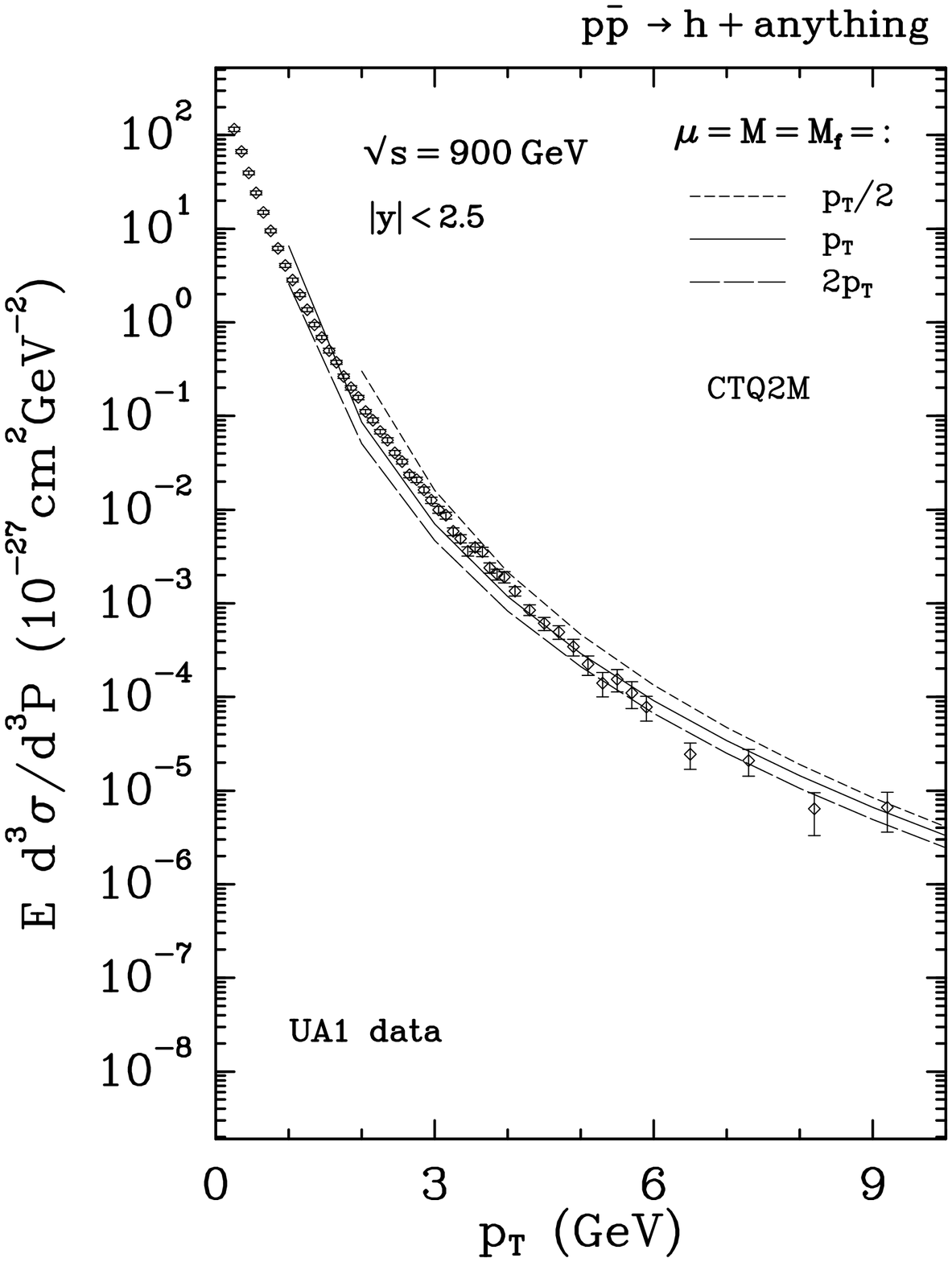}
{{\small {\bf Fig.~3.}
Same as in Fig. 1 for $\sqrt{s}=900\,$GeV }}
\end{figure}

The form of the coefficients $K_{a b \to c}$ is not unique. They
depend on the choice of finite corrections $f_{i,j}(x) $ and $d_{i,j}$
($i,j=q,g$) to structure and fragmentation functions. These
subtraction terms are accompanied by appropriate definitions of the
$F_a^p$ and $D_c^h$. We choose here the structure functions in the
$\overline{\rm MS}$ scheme. Specifically, we take the CTQ2M
($\overline{\rm MS}$) set of the proton distribution function
from\,[16] which gives a good fit to recent structure function
measurements at small $x$\,[17]. The starting scale of this set is
$M_0=2\,$GeV and therefore it can be used for prediction of the single
particle cross section at fairly small values of $\pT$. The $\as$ is
computed from the two-loop formula with
$\Lambda_{\overline{\rm MS}}=0.231\,$GeV as in the fit of the CTQ2M
($\overline{\rm MS}$) structure function.

As functions $D_c^h$, we employ the parametrizations by Binnewies et
al.\,[9]. They are given separately for the average of charged pions
and kaons, also in the $\overline{\rm MS}$ scheme. They are extracted
from the $\epem$ data on inclusive production of pions and kaons at
$Q=29\,$GeV and found in agreement with the majority of the $\epem$
data in the energy range between $5\,$GeV and $91\,$GeV. The
$\Lambda_{\overline{\rm MS}}=0.19\,$GeV chosen for these fits is
compatible with the $\Lambda_{\overline{MS}}$ in the proton structure
function. The starting energy $Q_0$, where fragmentation functions
have simple parametrizations, is chosen to be $Q_0=\sqrt{2}\,$GeV. The
$x$ and $Q^2$ dependence is then obtained by a NLO evolution.

The gluon fragmentation function into pions and kaons is poorely
determined in these fits to $\epem$ annihilation data. This is due to
the fact that the gluon participates in the process only at NLO. It
has an appreciable impact on the cross section only at very small
values of $x$, contributing mainly through the $Q^2$ evolution, due to
its coupling to the singlet combination of quarks. Therefore the gluon
parameters are correlated with the sea quark parameters and are not
very well constrained by $\epem$ data. To fix them independently, the
gluon fragmentation into charged particles is compared in\,[9] to the
three-jet data of the OPAL Collaboration at LEP\,[16].

In an earlier attempt\,[17] when this last comparison was not made, an
equally good description of all the other $\epem$ data used in\,[9]
could be obtained with weaker gluon fragmentation functions. To show
the gluon effect, we shall give for comparison, the inclusive charged
particle cross section with the weaker gluon distribution of\,[7] for
one of the $\pbp$ energies.

Further details of the fragmentations and their parametrization as a
function of $Q^2$ can be found in\,[9].

\begin{figure}[tb]
\epsfxsize=8.30cm
\epsfbox[61 131 534 752]{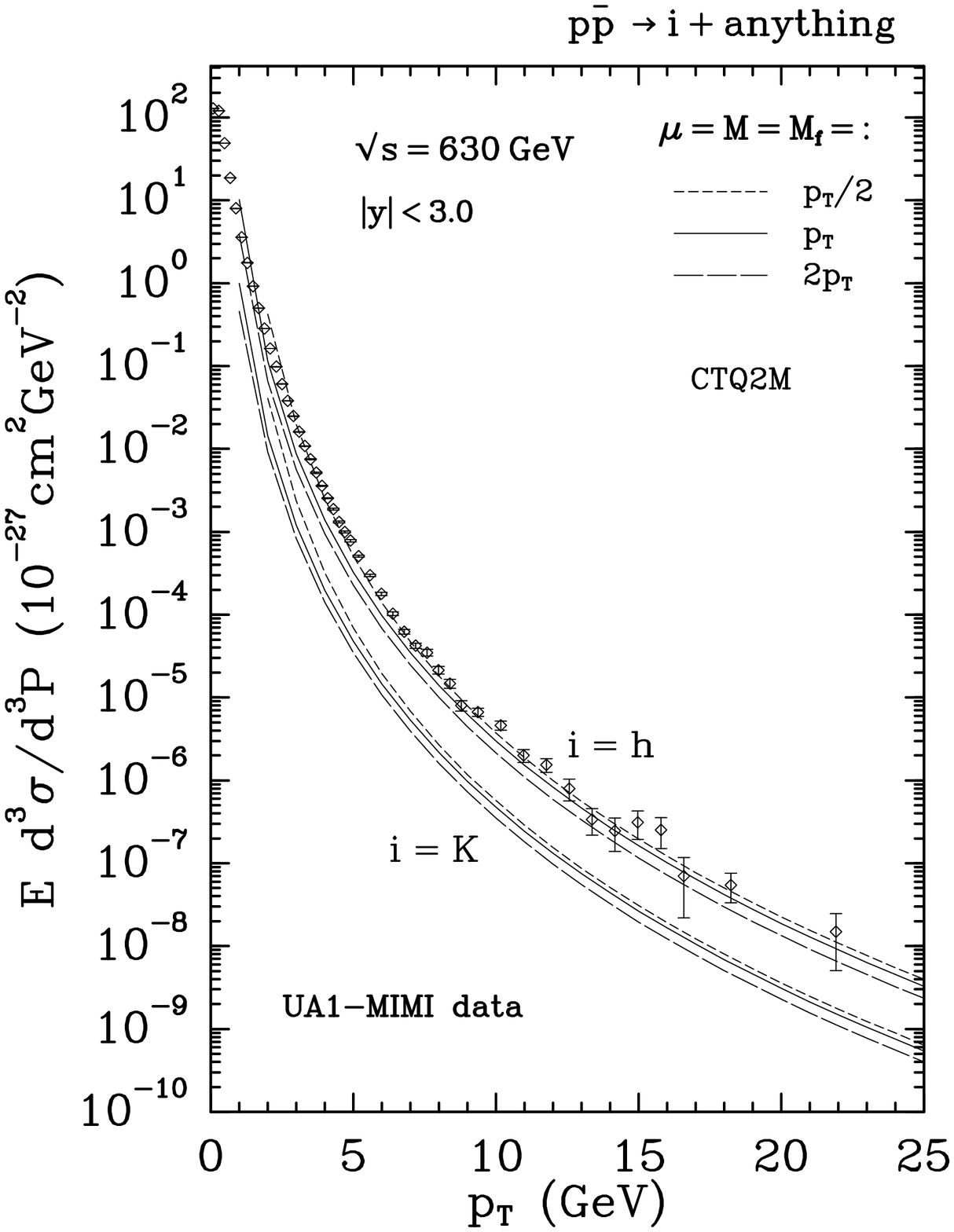}
{{\small {\bf Fig.~4.}
Inclusive cross section for production of hadrons $h \equiv
 h^+ + h^-$ and kaons, $K\equiv K^+ + K^-$ as a function of $\pT$ for
 $\sqrt{s}=630\,$GeV and rapidity
 range $\vert y \vert < 3.0$. The data obtained by
 UA1 MIMI Collaboration\,[6] for production of hadrons are also shown
}}
\end{figure}

\begin{figure}[tb]
\epsfxsize=8.0cm
\epsfbox[68 360 532 732]{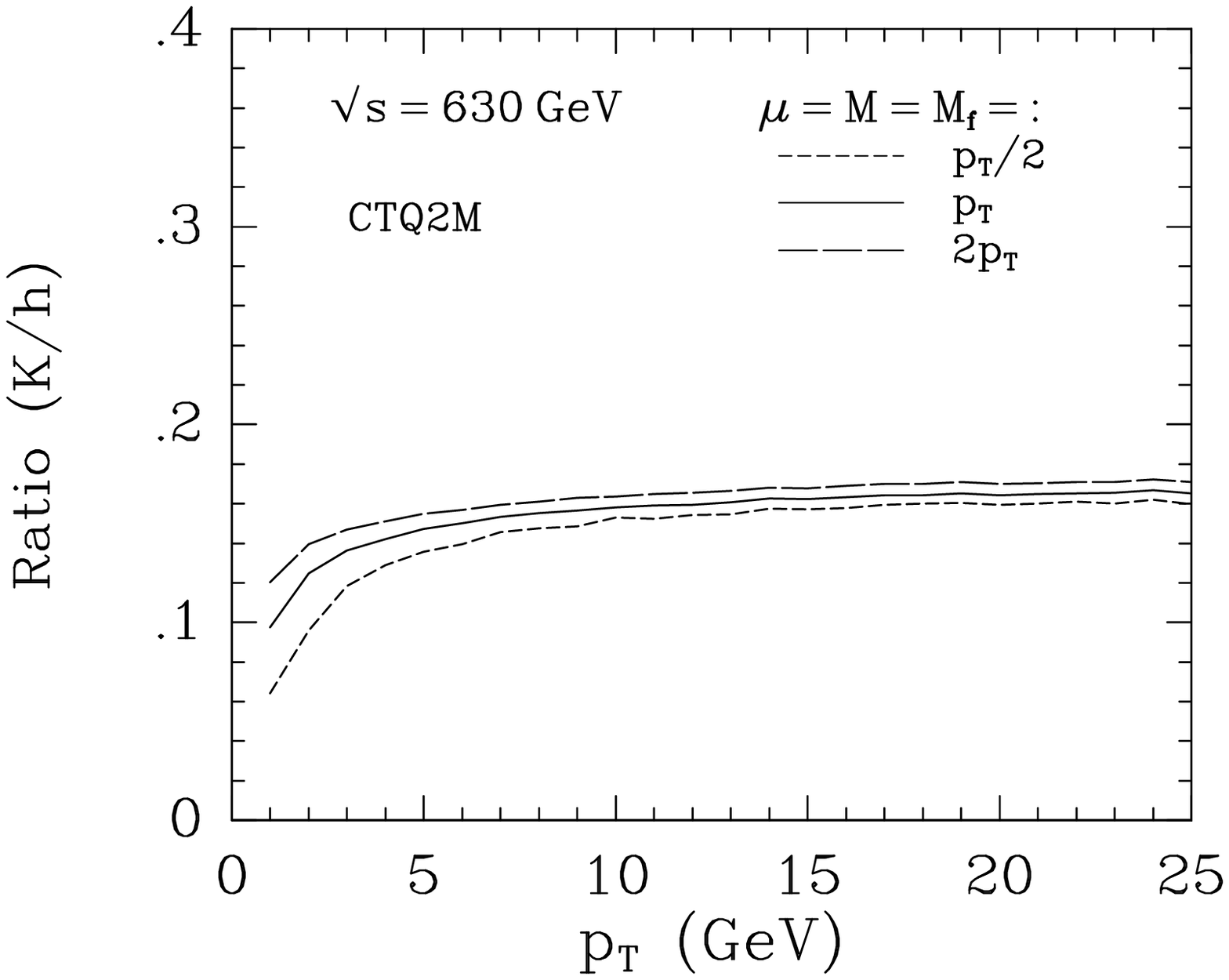}
{{\small {\bf Fig.~5.}
The ratio of inclusive cross sections for charged $K$,
$K\equiv K^++K^-$, over charged hadrons $h$, $h\equiv h^++h^-$, as a
function of $\pT$ for $\sqrt{s}=630\,$GeV and $\vert y \vert < 3.0$
 }}
\end{figure}

\section{Numerical results}
%
We are now in position to present our numerical results. We work
in the $\overline{\rm MS}$ scheme with $N_f=4$ active quark flavours.
We set the three scales $\mu$, $M$, $M_f$ equal and we vary them between
$\pT/2$ and $2\pT$. Unless otherwise specified the charged hadron $h$
is defined as $h\equiv (h^+ + h^-)/2 $, where $h^\pm$ sums over
$\pi^\pm$ and $K^\pm$.

In Figs. 1,2 and 3 we show the inclusive charged hadron cross section
for $p +\bar{p} \to h + X$ at $\sqrt{s}=200,500$ and $900\,$GeV. The
rapidity is averaged over the interval $-2.5 < y< 2.5$. The agreement
with the UA1 data\,[4] is best with scales equal to $\pT$, except at
$\sqrt{s}=200\,$GeV where the data lie somewhat nearer to the
prediction with scales equal to $\pT/2$. In agreement with the
experimental data, the theoretical curves show the expected increase
of the high $\pT$ tail between $\sqrt{s}=200\,$GeV and
$\sqrt{s}=900\,$GeV. According to\,[3], additional systematic errors
due to luminosity and acceptance corrections are small
($\pm15\%$). Therefore we can conclude that our predictions agree well
in shape and normalization with the data. The data with small $\pT$
have the smallest experimental errors. Unfortunately below
$\pT=2$--$3\,$GeV, our predictions cease to be valid: the soft
production mechanism takes over while the hard production mechanism
tends to give too fastly growing results.

The most recent and best experimental data come from the UA1 MIMI
Collaboration\,[6]. They have the smallest experimental errors and
extend earlier analysis of UA1 measurements to $\pT$ values up to
$25\,$GeV. These data are relative to production of charged hadrons
$h$, with $h\equiv h^+ + h^-$.  The theoretical predictions obtained
for the same definition of $h$ are plotted in Fig. $4$ for scales
$\pT/2, \pT$ and $2 \pT$, together with the data from\,[6]. The data
and the theoretical predictions are for $\sqrt{s}=630\,$GeV and are
averaged over $\vert y \vert < 3.0$. The agreement with the data from
the small up to the highest values of $\pT$ is excellent. The curve
with scale
$\pT/2$ fits the data best over the whole $\pT$ range above $3\,$GeV.

\begin{figure}[tb]
\epsfxsize=8.30cm
\epsfbox[61 131 534 752]{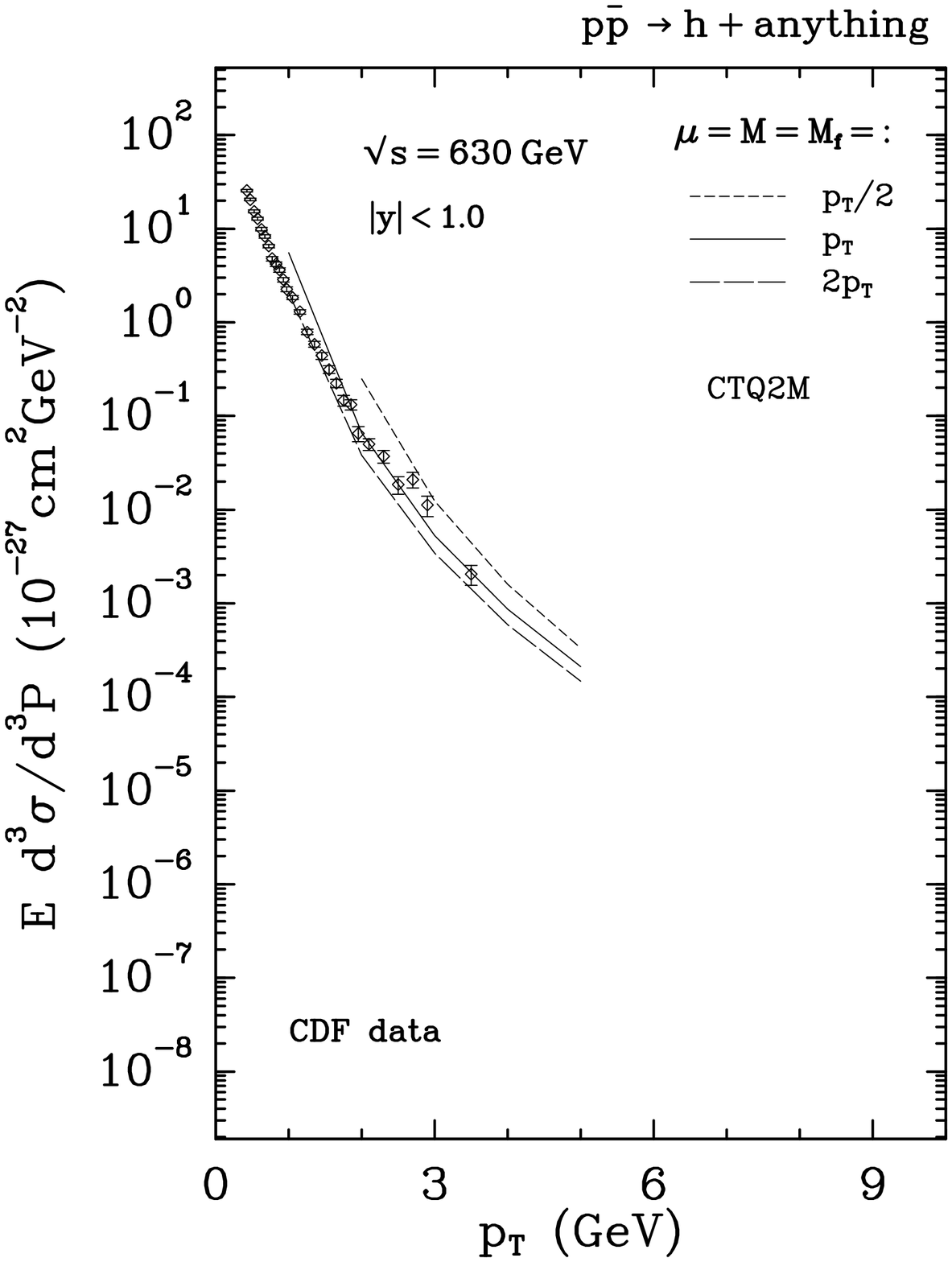}
{{\small {\bf Fig.~6.}
Same as in Fig. 1 for $\sqrt{s}=630\,$GeV and rapidity
 range $\vert y \vert < 1.0$. The theoretical results are compared with
 the data obtained by the CDF\,[5]
}}
\end{figure}
%
\begin{figure}[tb]
\epsfxsize=8.30cm
\epsfbox[61 131 534 752]{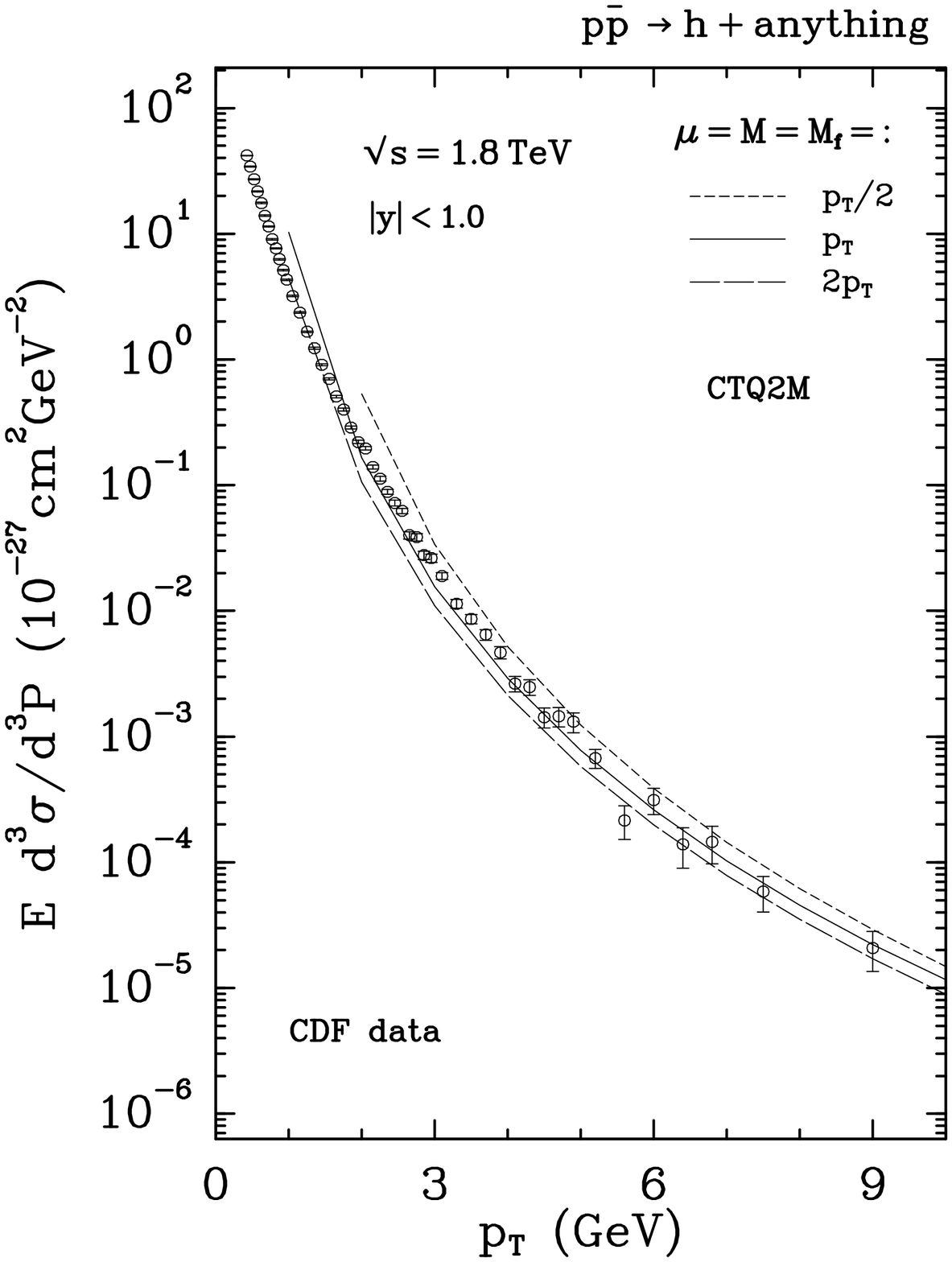}
{{\small {\bf Fig.~7.}
 Same as in Fig. 6 for $\sqrt{s}=1.8\,$TeV
}}
\end{figure}

In Fig. 4 we give also our prediction for $K$ production, in case such
data become available in the future. The ratio to the total charge
particle K/h is plotted in Fig. 5. Above $\pT=5\,$GeV this ratio is
approximately constant in $\pT$ with a slight increase towards larger
$\pT$. This ratio is approximately 0.15 and reflects the fact that the
fragmentation of quarks and gluons into $K$ mesons is much weaker than
the fragmentation into pions. Since $h$ is the sum of charged pions and
kaons, the inclusive charged pion cross section can be computed easily
from this ratio.

Similar conclusions can be drawn from the comparisons shown in Fig. 6
and 7 with CDF data\,[5] at $\sqrt{s}=630\,$GeV and
$\sqrt{s}=1.8\,$TeV, respectively. The definition of $h$ is here again
$h\equiv (h^+ + h^-)/2$. In both cases, the rapidity is averaged over
the interval $-1< y<1$. The data at both energies agree and are
predicted best with scales equal to $\pT$.

Our calculation seems to provide a rather good description of data
obtained in an energy range between $200\,$GeV and $1.8\,$TeV, with an
increase of energy by roughly one order of magnitude. Qualitatively
our results are independent of $\sqrt{s}$ at small $\pT$, i.e. below
$3\,$GeV, and increase with increasing energy in the large
$\pT$-range, as it is characteristic for a hard-scattering cross
section.

The degree of agreement between data and theoretical predictions,
however, is rather difficult to assess by inspection of the
logarithmic plots in Figs. 1-7, where the cross section drops rapidly
over several orders of magnitude. It is clear that at small $\pT$,
i.e. $\pT < 3\,$GeV, the theoretical curves deviate from the data. At
higher $\pT$, unfortunately, the data points have larger errors and
establishing the level of agreement becomes more problematic.

To make our comparison easier, we employ a three
parameter fit to the data of the form
$$
E {d^3\sigma \over d^3 p} =
    A \left( 1 + {\pT \over p_{{\rm T}_0}} \right)^{-n}\,.
\eqno(4)
$$
In\,[3] the $\sqrt{s}=500\,$GeV data of the UA1 Collaboration\,[3]
have been fitted to (4). This fit yields the parameters
$A= 408\pm 24 {\rm mb}/{\rm GeV}^2$,
$p_{{\rm T}_0} = 1.61 \pm 0.08\,$GeV and $n= 10.64\pm 0.31 $.
We divide our theoretical predictions for the three choices of scales
$\pT/2$, $\pT$, and $2 \pT$ as well as the experimental data by the
fit (4), taking for $A$, $p_{{\rm T}_0}$ and $n$ the central values.

\begin{figure}[h]
\epsfxsize=8.30cm
\epsfbox[62 360 522 732]{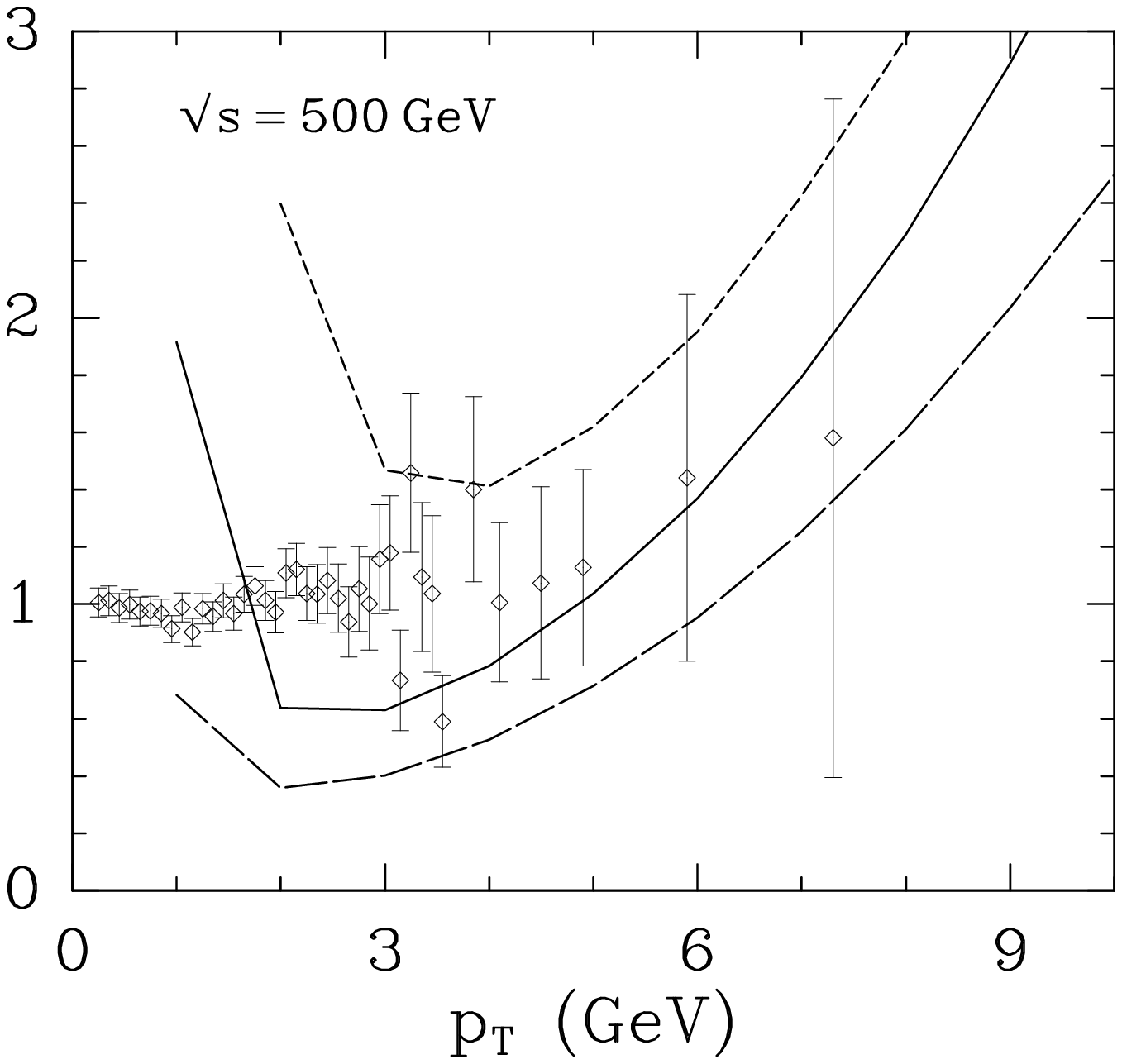}
{{\small {\bf Fig.~8.}
The ratio theoretical cross section over fitted ansatz for
 the cross section as given by (4). The short-dashed, solid and
 long-dashed lines correspond to the NLO predictions from Fig. 2 for
 scales $\mu=M=M_f$ set equal to $\pT/2$, $\pT$ and $2 \pT$, compared
 to the ratio of experimental data points to fitted ansatz (4) from
 Fig. 2 ($\sqrt{s}=500\,$GeV, $\vert y \vert < 2.5)$
}}
\vskip 0.3cm
\epsfxsize=8.30cm
\epsfbox[62 360 522 732]{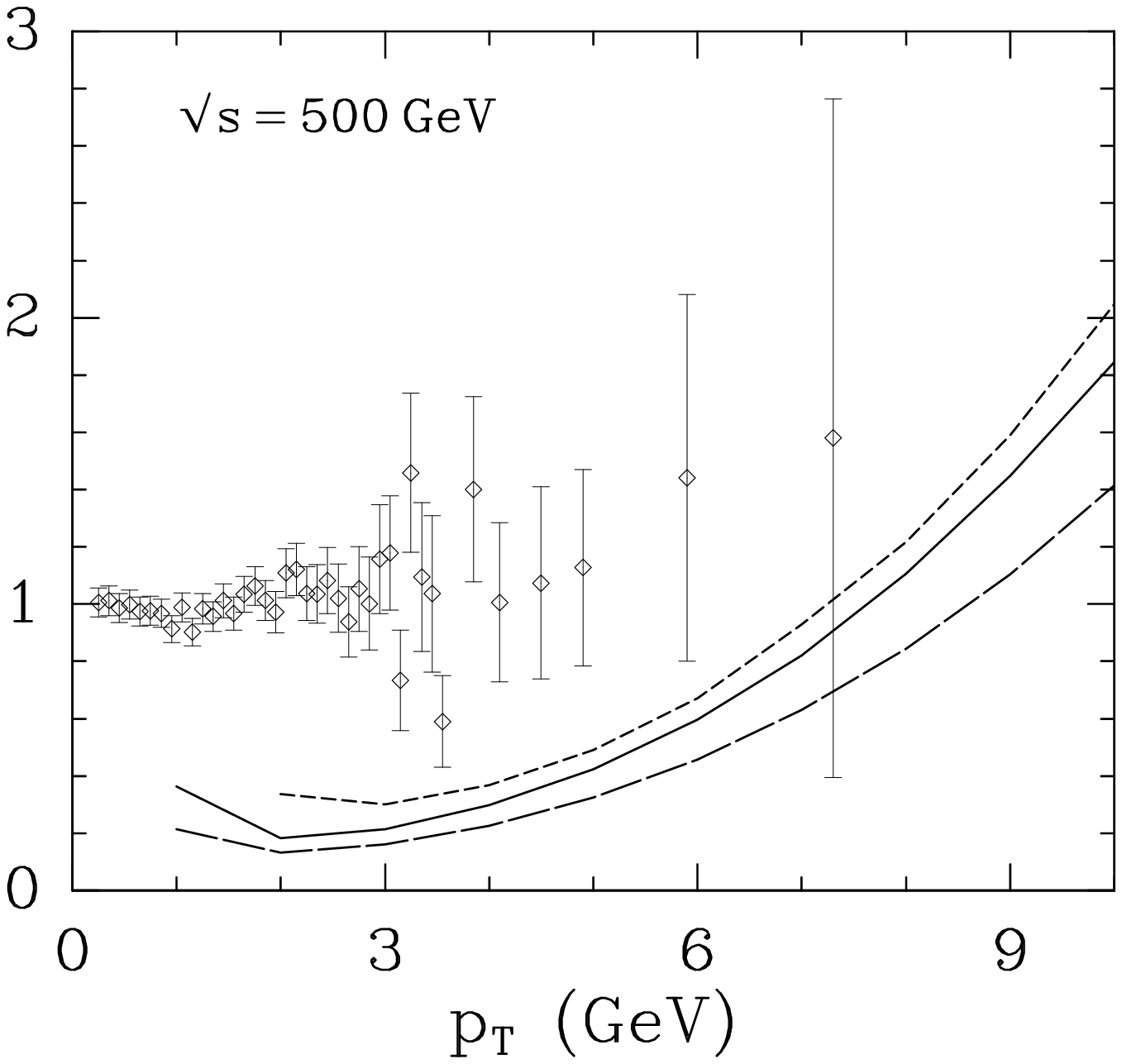}
{{\small {\bf Fig.~9.}
Same as Fig. 8 for the weaker gluon fragmentation given in\,[17]
}}
\end{figure}

The result is shown in Fig. 8.  It is clear that the fit reproduces
the data rather well for $\pT< 3\,$GeV, but it is less satisfactory
for $\pT$ around $3$--$4\,$GeV. It is compatible with the data for the
largest values of $\pT$ only thanks to the very large error of these
last data points. The ratio of the central value for the data over the
fit (4) seems to show an increase for increasing $\pT$. This is the
same behaviour shown by the theoretical predictions for which the
deviation from one becomes rather sharp at the largest values of $\pT$
here considered. It is more visible from this figure what was
previously said, i.e. that, overall, the choice of scales $\sim \pT$
in our calculation seems to give the best agreement with the
experimental data.

\begin{figure}[tb]
\epsfxsize=8.30cm
\epsfbox[61 131 534 752]{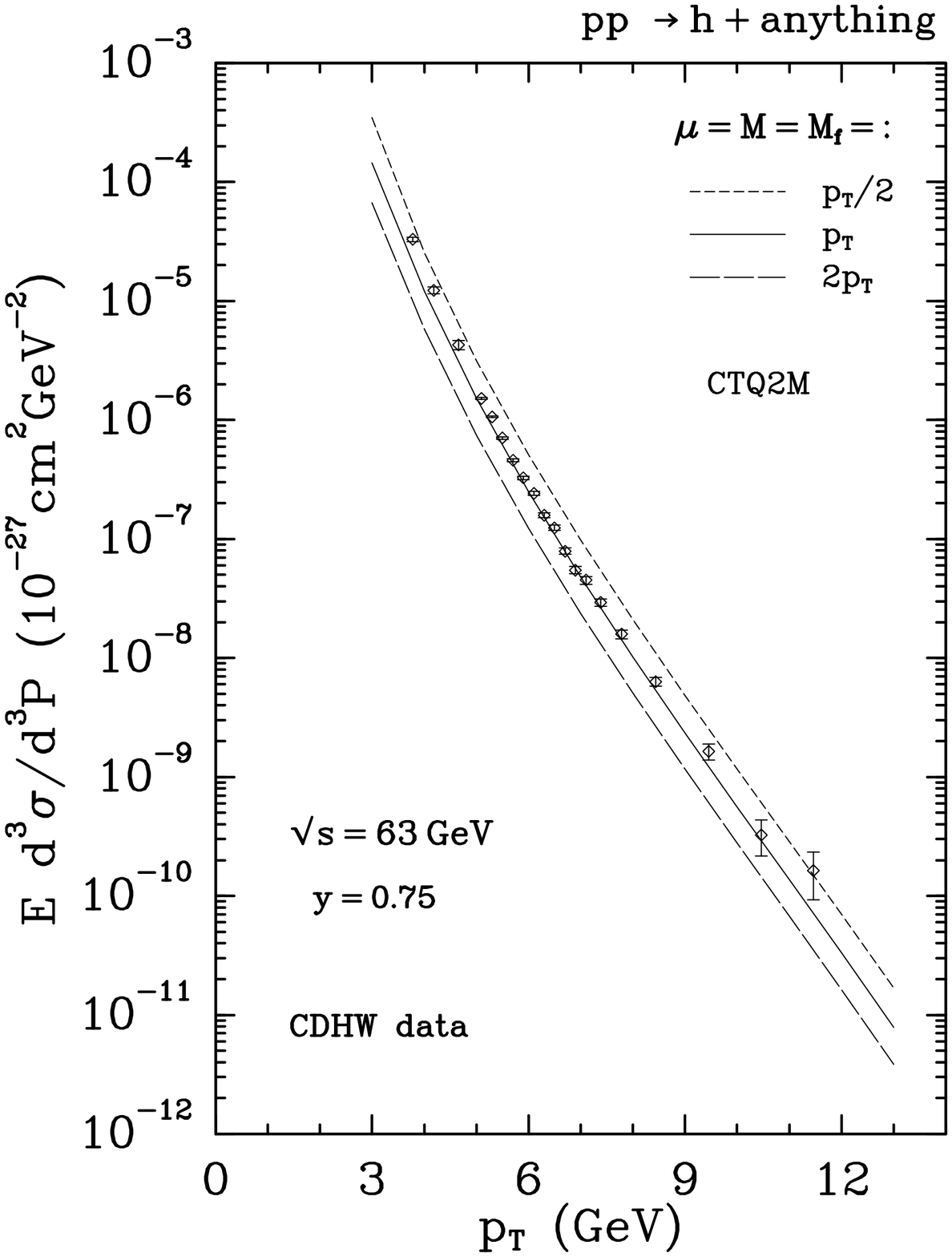}
{{\small {\bf Fig.~10.}
Inclusive cross section for production of hadrons
 $h \equiv h^+ + h^-$ as a function of $\pT$ for
 $\sqrt{s}=63\,$GeV and rapidity $ y = 0.75$. Also shown are the
 data obtained by the CDHW Collaboration\,[2] at the ISR $pp$ collider
}}
\end{figure}

It would be interesting to see in this same type of plots the effect of
different fragmentation functions. We do not attempt this here, but we
show in Fig. 9 the same ratios obtained by using the parametrization
of\,[17]. We remind here that this parametrization fits $\epem$ data
(except for the gluon fragmentation function from the OPAL
measurements\,[16]) equally well as the parametrization of\,[9]. The
gluon fragmentation function in\,[17], however, is weaker than the one
obtained in\,[9] where a fit to the OPAL data is also made. As one can
see the results shown in Fig. 9 disagree with the data. The
disagreement is stronger at smaller $\pT$: the theoretical predictions
are a factor $3$--$5$ away from the data for $\pT< 3\,$GeV. A
``strong'' gluon fragmentation to charged pions and kaons is therefore
needed to explain the inclusive charged particle cross sections in
$\pbp$ collisions.

So far all the inclusive particle production cross sections from
$\pbp$ are for the sum of all charged particles only. No separation
into particle species has been done. Such cross sections exist at
lower center of mass energies coming from the ISR $p p $
collider\,[2]. These data are for charged pion production at
$\sqrt{s}=63\,$GeV, $y\simeq 0.75$ and a range of $\pT$ between $3$
and $12\,$GeV. We have calculated this cross section with the same
input as for the $\pbp$ process (the $\bar{p}$ structure function in
(2) is obviously replaced by the structure function for $p$). The
result is shown in Fig. 10 together with the data from the CDHW
Collaboration\,[2]. The agreement with the theoretical curve with
scales equal to $\pT$ is excellent over the whole range of $\pT$.

We have calculated inclusive single-charged hadron cross sections in
full NLO, by using NLO structure functions, NLO fragmentation
functions for charged pions and kaons and NLO hard scattering cross
sections. Our results were compared with experimental data from the
CDHW, UA1, UA1-MIMI and CDF Collaborations. We found very good
agreement, in particular with the UA1-MIMI data which have the
smallest errors and extend over the largest $\pT$ range. The agreement
with the data is satisfactory for all center of mass energies between
$63\,$GeV and $1800\,$GeV in shape and absolute normalization, even in
the small $\pT$ region. We have demonstrated that only those
fragmentation functions with a large enough gluon contribution give a
satisfactory account of the collider data. The strength of this gluon
fragmentation
function agrees with the OPAL data in the three-jet region 
sensitive to it\,[16].

\noindent
{\bf Acknowledgments}

 The authors acknowledge the support from the
 Bundesministerium f\"ur Forschung und Technologie, Bonn, Germany,
 under contract 05~6~HH~93P(5),
 and of the EEC Program Human Capital and Mobility
 through Network Physics High Energy Colliders 
 CHRX-CT93-0357 (DG 12 COMA). F.B. acknowledges also the support
 from the Bundesministerium f\"ur Forschung und Technologie,
 Bonn, Germany, under contract 06~TM~743.

%

{\footnotesize
\begin{itemize}
\item[1.]
 J.F. Owens: Rev.\ Mod.\ Phys.\ 59 (1987) 465
\item[2.]
 D. Drijard et al., CDHW Coll.: Nucl.\ Phys.\ B208 (1982) 1
\item[3.]
 J.D. Dowell: in: 7th Topical Workshop on Proton-Antiproton
   Collider Physics, FNAL, 1988, R. Raja, A. Tollestrup,
   J. Yoh (eds.) Singapore: World Scientific 1988, p.~115;
 C. Albajar et al., UA1 Coll.: Nucl.\ Phys.\ B335 (1990) 261
\item[4.]
 M. Banner et al., UA2 Coll.: Phys.\ Lett.\ B122 (1983) 322 and
    Z. Phys.\ C27 (1985) 329
\item[5.]
 A. Para: in: 7th Topical Workshop on Proton-Antiproton
   Collider Physics, FNAL, 1988, R. Raja, A. Tollestrup,
   J. Yoh (eds.) Singapore: World Scientific 1988, p.~131;
 F. Abe et al., CDF Coll.: Phys.\ Rev.\ Lett.\ 61 (1988) 1819
\item[6.]
  G. Bocquet et al., UA1-MIMI Coll.: CERN preprint,
             CERN-PPE-94-47 (March 1994)
\item[7.]
  F.M. Borzumati, B.A. Kniehl, G. Kramer:
    Z. Phys.\ C57 (1993) 595
\item[8.]
  R. Baier, J. Engels, B. Petersson: Z. Phys.\ C2 (1979) 265;
  M. Anselmino, P. Kroll, E. Leader: Z. Phys.\ C18 (1983) 307
\item[9.]
 J. Binnewies, B.A. Kniehl, G. Kramer: DESY preprint, DESY 94-124
 (July 1994)
\item[10.]
 P. Chiappetta, M. Greco, J.Ph. Guillet, S. Rolli, M. Werlen:
   Nucl.\ Phys.\ B412 (1994) 3
\item[11.]
 M. Greco, S. Rolli: Z. Phys.\ C60 (1993) 169
\item[12.]
 M. Greco, S. Rolli, A. Vicini: Z. Phys.\ C65 (1995) 277
\item[13.]
 G. Marchesini, B.R. Webber: Nucl.\ Phys.\ B238 (1984) 1, ibid. B310
 (1988) 461
\item[14.]
 P.D. Acton et al., OPAL Coll.: Z. Phys. C58 (1993) 387
\item[15.]
  F. Aversa, P. Chiappetta, M. Greco, J.Ph.~Guillet:
   Phys.\ Lett.\ B210 (1988) 225; ibid.\ B211 (1988) 465;
   Nucl.\ Phys.\ B327 (1989) 105
\item[16.]
 H.L. Lai, J.F. Botts, J. Huston, J.G. Morfin, J.F. Owens,
 J. Qiu, W.-K. Tung, H. Weerts: Michigan State University
 Report No. MSU-HEP-41024,CTEQ-404, October 1994.
\item[17.]
 I. Abt et al., H1 Coll.: Nucl.\ Phys.\ B407 (1993) 515;
 M. Derrick et al., Zeus Coll.: Phys. Lett. B 316 (1993) 412
\item[18.]
 J. Binnewies: Diploma thesis (May 1994)
\end{itemize}
}
\end{document}